\documentclass[%
 reprint,
]{revtex4-2}

\usepackage{graphicx}% Include figure files
\usepackage{dcolumn}% Align table columns on decimal point
\usepackage{bm}% bold math
\usepackage{multirow}

\begin{document}

\title{On the decay properties of the neutron-deficient isotope $^{242}$Es}% Force line breaks with \\

		%%%% start author list %%%%
\author{J. Khuyagbaatar} \email{J.Khuyagbaatar@gsi.de}
\affiliation{GSI Helmholtzzentrum f\"ur Schwerionenforschung, 64291 Darmstadt, Germany}
\author{R.A. Cantemir}
\affiliation{GSI Helmholtzzentrum f\"ur Schwerionenforschung, 64291 Darmstadt, Germany}
\author{Ch.E. D\"ullmann}
\affiliation{GSI Helmholtzzentrum f\"ur Schwerionenforschung, 64291 Darmstadt, Germany}
\affiliation{Helmholtz-Institute Mainz, 55099 Mainz, Germany}
\affiliation{Johannes Gutenberg-Universit\"at Mainz, 55099 Mainz, Germany}
\author{ E. J\"ager}
\affiliation{GSI Helmholtzzentrum f\"ur Schwerionenforschung, 64291 Darmstadt, Germany}
\author{B. Kindler}
\affiliation{GSI Helmholtzzentrum f\"ur Schwerionenforschung, 64291 Darmstadt, Germany}
\author{J. Krier}
\affiliation{GSI Helmholtzzentrum f\"ur Schwerionenforschung, 64291 Darmstadt, Germany}
\author{N. Kurz}
\affiliation{GSI Helmholtzzentrum f\"ur Schwerionenforschung, 64291 Darmstadt, Germany}
\author{B. Lommel}
\affiliation{GSI Helmholtzzentrum f\"ur Schwerionenforschung, 64291 Darmstadt, Germany}
\author{ B. Schausten}
\affiliation{GSI Helmholtzzentrum f\"ur Schwerionenforschung, 64291 Darmstadt, Germany}
\author{A. Yakushev}
\affiliation{GSI Helmholtzzentrum f\"ur Schwerionenforschung, 64291 Darmstadt, Germany}
		%%%% end author list %%%%

\date{\today}% It is always \today, today,

\begin{abstract}
    
The radioactive decay properties of $^{242}$Es were studied with significantly improved statistics compared to available literature data. This isotope was produced in the 3n evaporation channel of the fusion reaction of $^{48}$Ca+$^{197}$Au. A half-life of 16.9(8)~s was deduced from 662 $\alpha$ decays of $^{242}$Es, resulting in an $\alpha$-decay branching of 41(3)\%. Twenty-six fission events with a half-life of 18.2$^{+4.5}_{-3.0}$~s were assigned to originate from the electron-capture delayed fission of $^{242}$Es. The probability for the electron-capture delayed fission was measured to be 0.015(4), which improves and resolves ambiguities in available experimental data. We discuss all known cases for electron-capture delayed fission in Es, Bk, and Am isotopes and compare experimental data with predictions from a recent semi-empirical model. A cross section of 27(3)~nb was measured for the production of $^{242}$Es.

\end{abstract}

\maketitle

%\tableofcontents

\section{Introduction}

The underlying nuclear shell structure of the heavy nucleus has a significant impact on its ground-state radioactive decay properties. In this regard, in odd-odd heavy nuclei, the shell stabilization effect is well pronounced compared to neighboring even-even, even-odd, and odd-even nuclei. This has been confirmed experimentally: the radioactive $\alpha$-decay and fission rates of the heavy nuclei are significantly reduced in nuclei with odd numbers of protons and neutrons. On the other hand, the nuclear structure of odd-odd nuclei is more complex compared to that in the neighboring ones. This is due to the coupling of neutron and proton orbitals occupied by a single nucleon. Thus, the exact configuration of the ground and of excited states of odd-odd heavy nuclei often remains unassigned experimentally. In turn, when $\alpha$ decay and fission are retarded in odd-odd heavy nuclei, the $\beta$ decay ($\beta ^{\pm}$ and electron capture, EC) may occur. This has been experimentally verified in nuclei with $Z$ up to 105. In the known isotopes of heavier elements, i.e., up to Og ($Z=118$), however, $\beta$ decay is not yet observed \cite{nndc}. 

Once odd-odd heavy nuclei undergo $\beta$ decay, the so-called $\beta$-delayed fission ($\beta$DF) may occur \cite{Berlovich69}. $\beta$DF is the process, where a part of the excited states that are populated in the even-even daughter nuclei can directly fission instead of feeding into the ground state via electromagnetic transitions. Accordingly, $\beta$DF provides access to low-energy fission where the macroscopic and/or microscopic effects of the underlying nuclear structure can be studied, even in the case of non-fissile nuclei (e.g., see review article \cite{andreyev13}). 

Most of the experimental data on $\beta$DF are from neutron-deficient nuclei, where electron-capture decay takes place.
The probability of electron-capture delayed fission (ECDF) is expressed as a ratio of DF to initial EC decays: $P_{\rm ECDF}=N_{\rm DF}/N_{\rm EC}$ \cite{andreyev13}. Recently, the $P_{\rm ECDF}$ was predicted to be quite large in EC-decaying superheavy nuclei (SHN, $Z>103$) \cite{Khu19d}. However, unambiguous evidence for the presence of ECDF in SHN has not yet been obtained, which is mainly due to the low production rates of SHN \cite{Khu17b}. On the other hand, a distinction between fission events from the DF and ground-state spontaneous fission is also quite a challenging issue. Thus, predictions from Ref.~\cite{Khu19d}, i.e., large probabilities of ECDF in SHN, still need to be verified. 

At the same time, the predictions from Ref.\cite{Khu19d} can be tested with heavy nuclei, which can be produced with reasonably large statistics. As one example, recently, the ECDF was searched for in the new isotope $^{244}$Md \cite{Khu20b}. However, the EC decay of this isotope was not observed due to insufficient statistics, and only an upper limit for its branching ratio was deduced. 

Presently, the $P_{\rm ECDF}$ in Am, Bk, and Es isotopes represents the best systematic data set, which is the main source for the theoretical description of the ECDF process in Ref.~\cite{Khu19d}. However, in some cases, the available experimental data (e.g., $P_{\rm ECDF}$) on the ECDF still needs to be improved, as recently made in $^{188}$Bi \cite{Andel20}. One such case is $^{242}$Es, which has both $\alpha$ and EC decay branches. Its $P_{\rm ECDF}$ has been measured in several independent experiments, where $^{242}$Es was produced in different reactions \cite{Hingmann85,Lazarev94,shaughnessy00,antalic10}. In the first ECDF-experiment \cite{Hingmann85}, the fusion-evaporation reaction of $^{40}$Ar~+~$^{205}$Tl was used, and three fission events resulting in $P_{\rm ECDF}=0.014(8)$ were measured. In Ref.~\cite{antalic10}, $^{242}$Es was produced in the $\alpha$ decay of $^{246}$Md. The same number, i.e., three fission events were measured resulting in $P_{\rm ECDF}=0.013^{+0.012}_{-0.007}$. These experimental values measured with the so-called focal plane implantation technique are in agreement within their large statistical uncertainties. Meanwhile, the ECDF of $^{242}$Es was studied with significantly larger statistics (48 fission events) in Ref.~\cite{shaughnessy00}, where the $^{14}$N~+~$^{233}$U reaction was used. However, the ECDF probability was only 0.006(2), which is about two times smaller than the aforementioned two values. In fact, this experiment did not implant the $^{242}$Es into a silicon detector as the other two, but instead the $^{242}$Es was transported to a fission measurement station with an aerosol-gas jet. Accordingly, only the DF events from $^{242}$Es were measured. 

Therefore, the present work is aimed at measuring the ECDF probability of $^{242}$Es precisely by detecting all its $\alpha$, EC, and fission branches simultaneously. The $^{242}$Es isotope was produced in the scarcely known fusion-evaporation reaction of $^{48}$Ca~+~$^{197}$Au \cite{Bris19}. 

\section{Experimental setup}

The experiment was performed at the gas-filled TransActinide Separator and Chemistry Apparatus (TASCA) at GSI \cite{SemB08a}. A pulsed (5~ms-long pulses, $\approx$50/s repetition rate) $^{48}$Ca$^{10+}$ beam was accelerated by the Universal Linear Accelerator UNILAC to 218.9~MeV energy. In the present experiment, mostly 5~Hz of the pulsed beam was used. A gold target with an average thickness of about 0.50~mg/cm$^{2}$ was mounted on a wheel that rotated synchronously with the beam macro-structure \cite{JaeB14a}. The beam energy in the middle of the target was 215.9~MeV, which corresponds to an excitation energy of $\approx$32~MeV of the compound nucleus $^{245}$Es$^{\ast}$ formed in the $^{48}$Ca~+~$^{197}$Au reaction \cite{Rei81}. 

For separation and collection of evaporation residues (ERs) in the focal plane detector, TASCA was operated with helium gas at 0.8\,mbar pressure and with a magnetic rigidity ($B\rho$) of 2.0~Tm \cite{Khu12a,Khu13b}. The efficiency of TASCA to guide ERs to the focal plane implantation detector (stop detector) was estimated to be 60\% \cite{Gre13a,Khu12a}. A multi-wire proportional counter (MWPC) was mounted in front of the stop detector and used to discriminate between events passing through TASCA and events originating from the radioactive decays of implanted ERs. Double-sided silicon strip detectors were used as a stop detector, and they comprised a total of 144 vertical ($X$) and 48 horizontal ($Y$) strips on the front and back sides, respectively. The stop detector was cooled by liquid alcohol circulating at a temperature of -20$^{\circ}$\,C. Energy resolutions (FWHM) of both $X$ and $Y$ strips were about $40$\,keV for measurements of external $\alpha$ particles with 5.8~MeV. Energy calibrations of detectors were made using $\alpha$ decays of the implanted nuclei produced in the $^{48}$Ca\,+\,$^{176}$Yb reaction. The efficiency for full-energy $\alpha$-particle detection in the stop detector was $\approx$55\%.

Signals from $X$ and $Y$ strips were shaped and amplified with different gains to provide two energy ranges up to about 20\,MeV and 200\,MeV, respectively. After amplification, all signals were digitized by 100\,MHz-sampling FEBEX4 14-bit ADCs developed by the GSI experiment electronics department \cite{Febex4}. The shape of each signal was stored in a 30~$\mu$s-long trace with 5~$\mu$s pre-trigger time.
Details on the data acquisition and detection systems of TASCA can be found in Refs.~\cite{Khu19b,Khu21b}

The average beam intensities on the target were about 100~pnA resulting in a total counting rate of about 60~Hz in the stop detector.

\begin{figure}[t]
	\vspace*{-0mm}
	\centering
	\hspace*{0mm}
	\resizebox{0.47\textwidth}{!}{\includegraphics[angle=0]{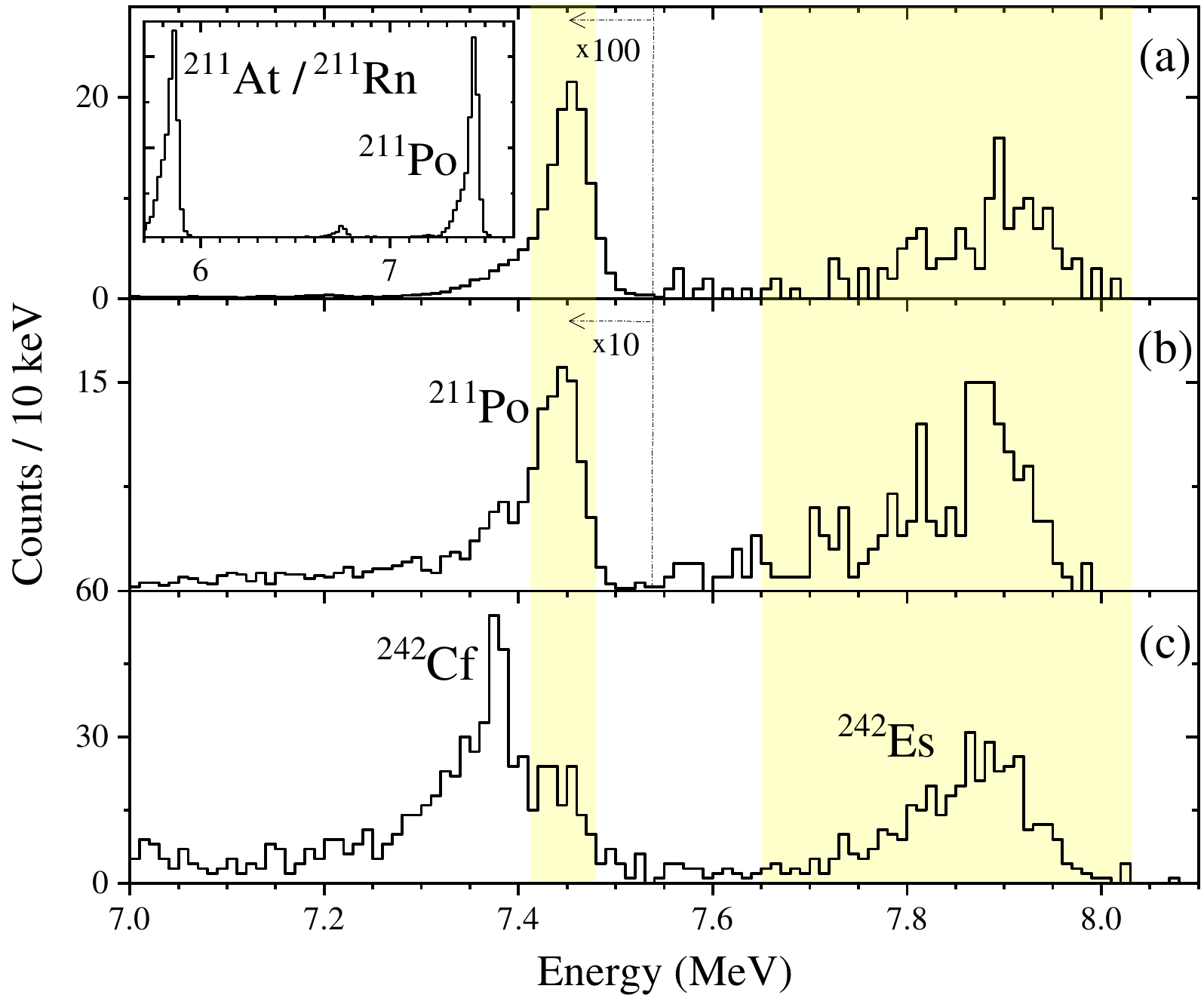}}
	\vspace*{-4mm}
	\caption{(Color online) Energy spectra of the stop detector ($X$-strips) measured during the beam-off periods from the $^{48}$Ca\,+\,$^{197}$Au reaction in run~1 (a), run~2 (b) and run~3 (c). In the inset of (a), the energy spectrum measured in run~1 is shown in a broader energy range. Isotopic assignments of peaks are marked. See text for details.
	}
	\label{spectra}
\end{figure}

\section{Experimental results and discussions} 

The energy spectra measured from $X$-strips of the stop detector during the beam-off periods from three sets of irradiation (runs), i.e., $^{48}$Ca~+~$^{197}$Au are shown in Fig.~\ref{spectra}. These measurements were performed in the same experimental campaign but with time intervals of about four and two days between the runs 1-2 and 2-3, respectively. In all runs, the background $\alpha$ events originating from $^{211}$Po, which has a half-life of $\approx$0.5~s \cite{nndc} were detected. During the experiment, this isotope was produced as an EC-decay product of the long-lived $^{211}$At (7.2~h, \cite{nndc}) and $^{211}$Rn (14.6~h, \cite{nndc}). In the inset of Fig.~\ref{spectra}(a), the spectrum from run~1 is shown in a wider energy range. The peak at an energy of $\approx$5.86~MeV corresponds mainly to $\alpha$ decays of $^{211}$At, with some contributions from $^{211}$Rn. 
In turn, these long-lived isotopes were produced in the $\alpha$-decay chains of Ra-Th, which were directly produced in the fusion-evaporation reaction $^{48}$Ca~+~$^{176}$Yb. As mentioned above, this is a typical reaction, which we use frequently to calibrate the stop detector. This reaction was measured prior to the runs~1 and 2, thus, the corresponding spectra have a large contamination from $^{211}$Po. In between the runs~2 and 3, no calibration reaction was used. This resulted in only small amounts of long-lived contaminants, which led to a small background from $^{211}$Po as seen in Fig.~\ref{spectra}(c).

In all three measurements, a well-pronounced but broadly distributed peak-like structure in the energy range of 7.65-8.05~MeV is observed.
The energy distribution of these events is similar to that previously measured for $^{242}$Es. The latter exhibits an unresolved yet complex $\alpha$-decay fine structure \cite{antalic10}. A broad energy distribution of the $\alpha$ events indicates the population of low-lying excited states, which strongly decay by internal conversion. In \cite{antalic10}, the $\alpha$-particle energies have predominantly been observed in the range of 7.78-7.96~MeV and a group of events with energies of 8.025(20)~MeV. Presently observed events are in agreement with these results despite a broader range of $\alpha$-particle energies, which we attribute to the relatively poor energy resolution of the stop detector compared to the one used in Ref.~\cite{antalic10}. Accordingly, all events in the energy range of 7.65-8.05~MeV can be attributed to the $\alpha$ decay of $^{242}$Es. We examined the traces of $\alpha$ events for the presence of any small-energy signal from delayed conversion electron \cite{Khu19c}, which may originate from a spin isomeric state in $^{238}$Bk. We did not find any such delayed signals. Note that signals from the $\alpha$-particle and the conversion electron cannot be resolved when the time between them is shorter than $\approx$100~ns \cite{Khu22}.

At the same time, we do not exclude contamination due to the $\alpha$ decay from $^{243}$Es, which can be produced in the 2n channel of the fusion-evaporation reaction of $^{48}$Ca~+~$^{197}$Au. This isotope emits $\alpha$ particles with an energy of $\approx$7.89~MeV and has an $\alpha$-decay branching of 61\% \cite{nndc}. Presently, $\alpha$-particles from $^{243}$Es are impossible to be distinguished from the $\alpha$ decay of $^{242}$Es. By using the statistical code HIVAP for the prediction of cross sections of the fusion-evaporation reaction \cite{Rei81}, we estimated that about 10\% of the $\alpha$ events in the range of 7.66-8.05~MeV should originate from the decay of $^{243}$Es. In the following analysis and discussion, we will show that this is reasonable and that it does not have a significant statistical effect on the results for $^{242}$Es. 

\begin{figure}[t]
	\vspace*{2mm}
	\centering
	\hspace*{0mm}
	\resizebox{0.43\textwidth}{!}{\includegraphics[angle=0]{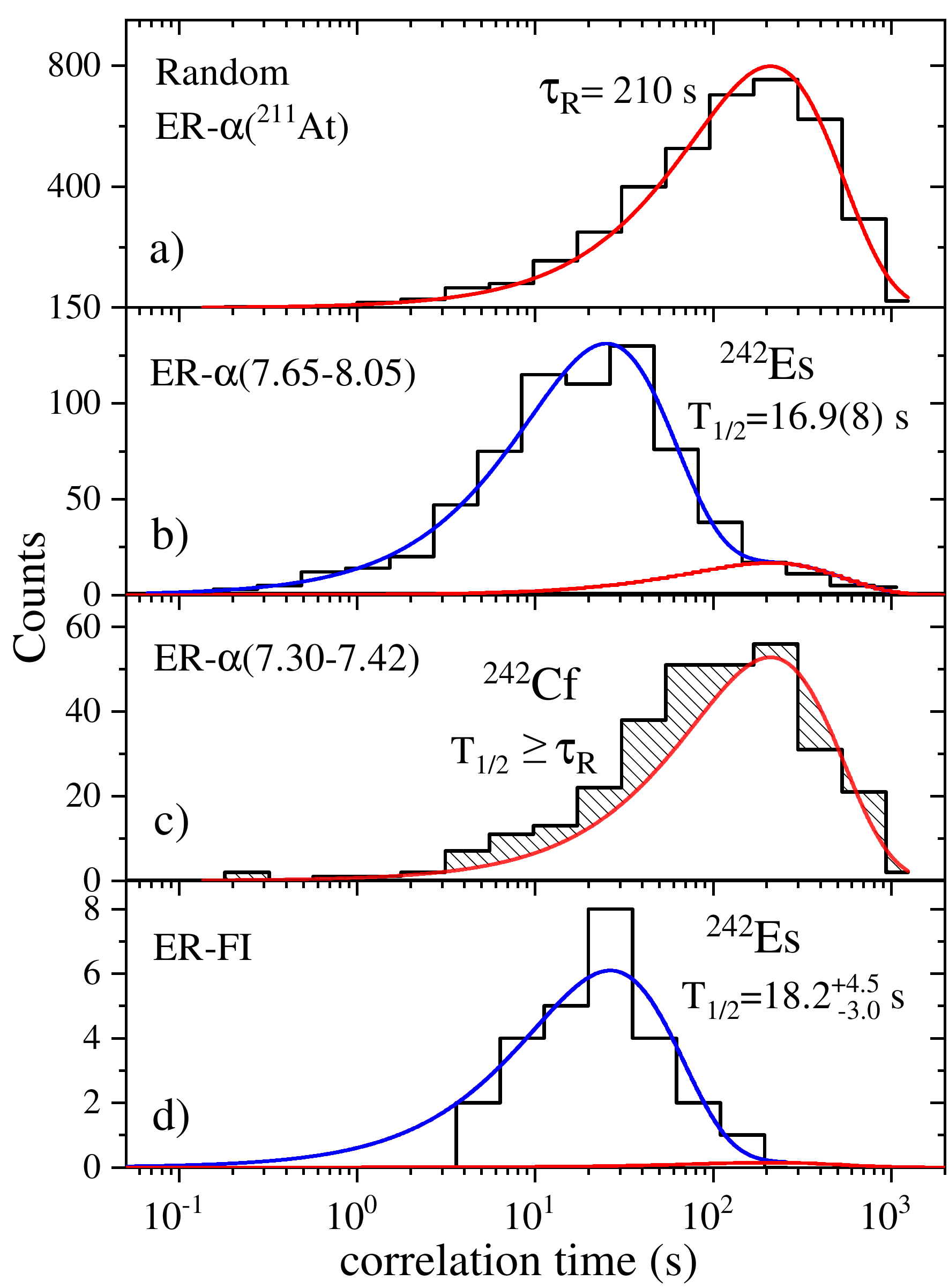}}
	\vspace*{-4mm}
	\caption{(Color online) The time distributions of $\alpha$ and fission events correlated with preceding implantation signals. The random correlation (a), $\alpha$ events assigned to the decay of $^{242}$Es (b), $\alpha$ events assigned to the decay of $^{242}$Cf and the fission events assigned to the  ECDF of $^{242}$Es. Fits of the experimental data by the time-density distribution function according to Ref.~\cite{SchS84a} are shown. See text for details.
	}
	\label{lifetimes}
\end{figure}

The half-life of $^{242}$Es can be deduced from the spatial and time correlation analysis between the implantation signals, i.e., ERs and $\alpha$ events in the range of 7.65-8.05~MeV. But, first, the random rate of the ER-like events was analyzed. For this, we used the $\alpha$ events from the peak corresponding to $^{211}$At (see inset of Fig.~\ref{spectra}(a)). Since this long-lived isotope was produced as the daughter of Ra-Th, no ERs should be found. Thus, any ER--$\alpha$($^{211}$At) event should be a random correlation. ER-like events were required to have an energy in the range of 5-25~MeV, and to be anticoincident with any signal from the MWPC. The time distribution of $\alpha$($^{211}$At) events correlated with the only one and nearest preceding ER-like signals is shown in Fig.~\ref{lifetimes}(a). The correlation search time was $2000$~s. The time distribution can be described by the time-density distribution function given in Ref.~\cite{SchS84a}. An average random correlation time of $\tau_{R}=210$~s is extracted. This random ER correlation rate should be considered when ERs leading to the $\alpha$ and fission events of the $^{242}$Es and $^{242}$Cf will be searched.

The time distribution of ER--$\alpha$(7.65-8.05~MeV) events is shown in Fig.~\ref{lifetimes}(b). Again, we considered only the nearest preceding ER-like signal for each $\alpha$ event. A well-pronounced non-random peak is observed. Meanwhile, in the longer time range, the events with a random origin are visible. Hence, the measured time distribution was fit with the two time-density distribution functions, and the result is shown in  Fig.~\ref{lifetimes}(b). The non-random component is well-described by the radioactive decay with a half-life of 16.9(8)~s. This value is in fine agreement with the previously measured value of 17.8(16)~s for $^{242}$Es \cite{antalic10}. At the same time, both of these values are slightly larger than the 11(3)~s measured in the gas-jet transport technique in Ref.~\cite{shaughnessy00}. Nevertheless, the $\alpha$ events with energies in the range of 7.65-8.05~MeV were unambiguously attributed to the decay of $^{242}$Es. We did not observe any small-energy signal in the trace of any ER correlated with an $\alpha$(7.65-8.05~MeV) event. Accordingly, within the accumulated statistics no isomeric state with $\mu$s half-life in $^{242}$Es was observed.

For the identification of the EC decay of $^{242}$Es, we should consider both $\alpha$ decay of the EC daughter $^{242}$Cf (61\%, \cite{nndc}) and fission from the excited $^{242}$Cf.

Alpha-particles from $^{242}$Cf with an energy of 7.39~MeV \cite{nndc} should be detected in the region of 7.3-7.4~MeV. However, the energy spectra measured during the runs~1 and 2, as seen in Fig.~\ref{spectra}, show a large background peak at 7.45~MeV. Thus, this impedes the identification of $\alpha$ decay originating from $^{242}$Cf. The origin of these background-contamination $\alpha$ events was discussed above. In the run~3, though, the peak at $\approx$7.39~MeV, which can be attributed to the decay of $^{242}$Cf becomes visible. Thus, the ERs were searched for these $\alpha$ events with energies of 7.30-7.42~MeV and the result is shown in Fig.~\ref{lifetimes}(c). The time distribution, however, matches the expected random one. It is worth noting that the shape of this time distribution is not perfectly described by the random one, which may indicate the presence of true ER--$\alpha$ correlations originating from $^{242}$Cf. This isotope is known to have a half-life of 3.4~min (204~s), which is indeed comparable with the present random correlation time. Therefore, presently, we cannot determine the half-life of $^{242}$Cf. On the other hand, by using the data from run~3, we can estimate the EC-decay branching of $^{242}$Es as will be shown below.

We looked at events with energies between 50 and 200 MeV to find the EC-decay branching of $^{242}$Es that led to fission from an excited state. We select only those events detected during the beam-off periods and without the MWPC signal. A total of 26 (five in run~1, eight in run~2, and thirteen in run~3) events, which can be attributed to being fission-like (FI events), were observed (see Fig.~\ref{spectra}(d-f)). These FI events do not originate from the spontaneous fission of $^{242}$Cf, which has a negligibly low branching ($\leq$0.014\%, \cite{nndc}). All these FI events were found to be correlated with ERs, and their time distribution is shown in Fig.~\ref{lifetimes}c. These events have non-random origins and are well described by the radioactive decay curve with a half-life of 18.2$^{+4.5}_{-3.0}$~s. This value is in fine agreement with the above-determined half-life of 16.9(8)~s for $^{242}$Es. Therefore, these FI events are assigned to originate from the fission of excited states in $^{242}$Cf produced in the EC decay of $^{242}$Es. 

We used the data from the run~3 to extract final values for the $\alpha$-decay and EC-decay branches of $^{242}$Es. We take the events in the region of 7.30-7.42~MeV as belonging to the decay of $^{242}$Cf. This resulted in 337 $\alpha$ events, which mostly form the peak at $\approx$7.38~MeV. By taking into account the $\alpha$-decay branching of $^{242}$Cf (61(3)\%, \cite{nndc}), and 384 ER--$\alpha$ events (correlation time less than 200~s) of $^{242}$Es, we extracted the EC-decay branching of 59(3)~\% for $^{242}$Es. This value is in fine agreement with the literature value of 57(3)~\% \cite{antalic10}. The FI events (13) were not taken into account, and they do not alter the deduced mean value for the EC-decay branching. 

For extracting a final value for $P_{\rm ECDF}$ of $^{242}$Es, we used the data measured from all runs~1-3. The total numbers of 662 ER--$\alpha$ and 26 ER--FI events with correlation times of less than 200~s were attributed to $^{242}$Es and $^{242}$Cf$^{\ast}$, respectively. By using the presently measured EC-decay branching of 59(3)~\% for $^{242}$Es, the $P_{\rm ECDF}$ of $^{242}$Es was deduced to be 0.015(4). This value is significantly improved compared to all previously measured values. 

Based on the aforementioned number of $\alpha$ events and presently deduced $\alpha$-decay branching for $^{242}$Es, we calculated its production cross section of 27(3)~nb in the 3n channel of the $^{48}$Ca~+~$^{197}$Au reaction. 

Presently, we could not unambiguously identify the ER--$\alpha$($^{242}$Es)--FI chain, which would belong to the ECDF of $^{238}$Bk. We found ERs for all FI events, thus, no FI event was left without identification. However, we should note that the random rate of ERs was relatively high, which could result in one random event in Fig.~\ref{lifetimes}(d). On the other hand, we expected to observe two random events forming the ER--$\alpha$($^{242}$Es)--FI sequences with a time correlation of up to 100~s between the last members. In the analysis, we found two such triple-correlation events, as expected, and they were attributed to being random. Based on this, we estimated an upper limit of 1.6$\times 10^{-3}$ for the ECDF of $^{238}$Bk. This limit is in line with the literature value of 4.8(20)$\times 10^{-4}$ given in \cite{Kreek94}. Eventually, this value was not directly measured in \cite{Kreek94}, but it was estimated on the basis of the calculated cross section of $^{238}$Bk in the $^4$He~+~$^{241}$Am reaction. The presently measured decay properties for $^{242}$Es and $^{238}$Bk are summarized and compared with the literature data in Table~\ref{rates}.

\begin{figure*}[t]
	\vspace*{-0mm}
	\centering
	\hspace*{-2mm}
	\resizebox{0.8\textwidth}{!}{\includegraphics[angle=0]{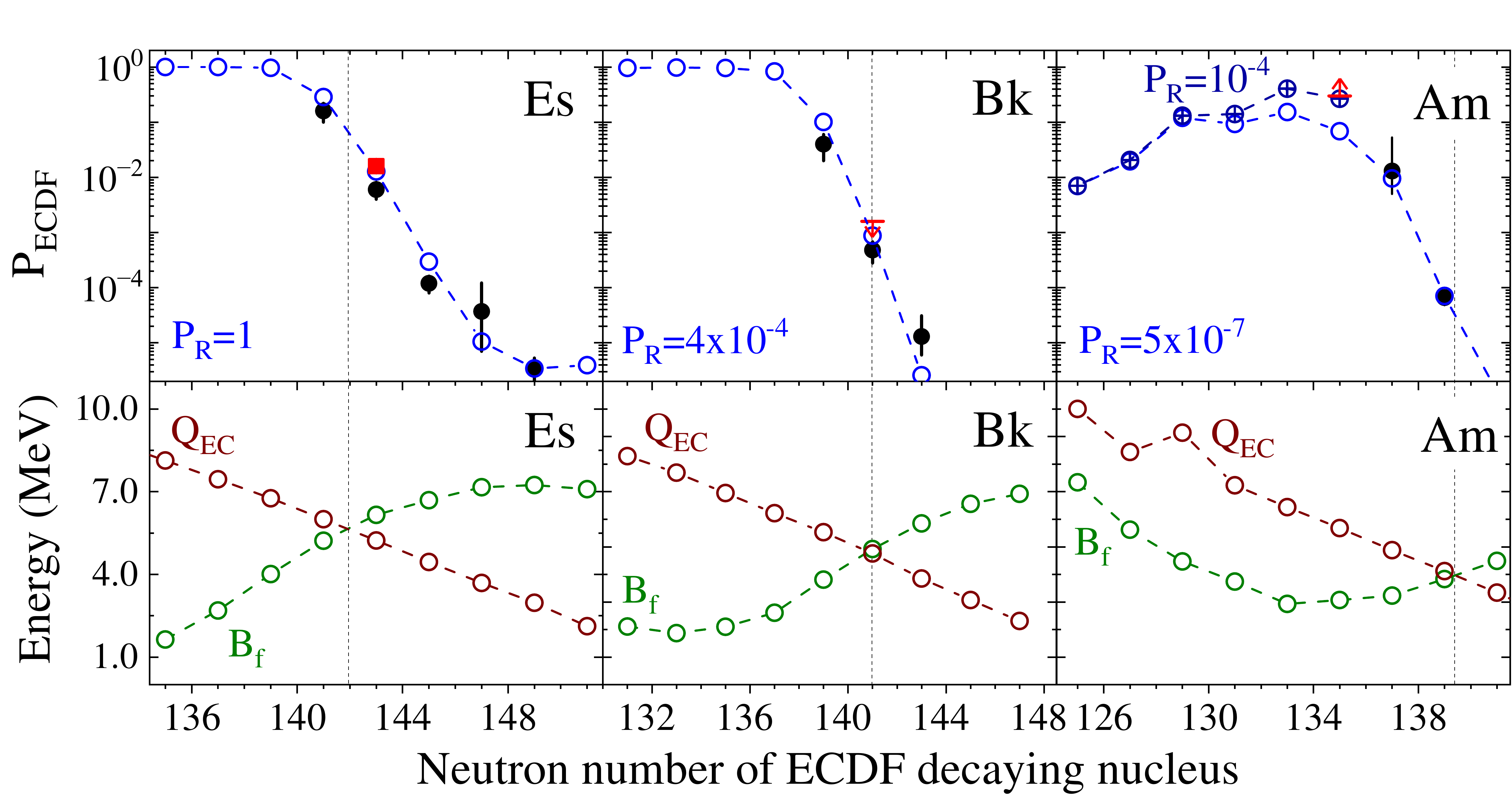}}
	\vspace*{-2mm}
	\caption{(Color online) Upper panels: probabilities of ECDF in Es, Bk and Am isotopes. Solid and open circles mark the experimental \cite{andreyev13,Konki2017265} and calculated \cite{Khu19d} $P_{\rm ECDF}$, respectively. A retardation factor, $P_{\rm R}$ used for the description of the delayed fission probabilities according to Ref.~\cite{Khu19d} and corresponding to calculated values (open circles) is given for each element. The presently measured value for $^{242}$Es is shown by a solid square. Arrows indicate the present upper limit for $^{238}$Bk and the recently reported lower limit for $^{230}$Am \cite{Wilson17}. Lower panels: theoretical $B_{\rm f}$ and $Q_{\rm EC}$ values taken from Refs.~\cite{Moll09} and \cite{Moll95} are shown as a function of the neutron number of the EC-decaying odd-odd nucleus. Dashed vertical lines mark the crossing of the $Q_{\rm EC}$ and $B_{\rm f}$ curves. See texts for details.
	}
	\label{PECDF}
\end{figure*}

\begin{table}[thb]
	\caption{The decay properties of Es and Bk isotopes measured in the present work are compared with the most recent literature data.}
	\begin{ruledtabular}
		\begin{tabular}{llcc}
			\noalign{\smallskip} 
		
			& & this work & lit.  \\\noalign{\smallskip} 
			
\multirow{4}{*}{$^{242}$Es} & $b_{\alpha}$ (\%) & 41(3) & 43(3) \cite{antalic10}
 \\\noalign{\smallskip} 
			& $T_{1/2}$ (s) & 16.9(8) & 17.8(16) \cite{antalic10}  \\\noalign{\smallskip}
			&  & & 11(3) \cite{shaughnessy00} 
				 \\\noalign{\smallskip} 
			& $P_{\rm ECDF}$ & 0.015(4) & $0.013^{+0.012}_{-0.007}$ \cite{antalic10}  \\\noalign{\smallskip}
 			&  & & 0.014(8) \cite{Hingmann85}
			\\\noalign{\smallskip} 
&  & & 0.006(2) \cite{shaughnessy00} 
\\\noalign{\smallskip} 				 			
			$^{238}$Bk	& $P_{\rm ECDF}$ & $<1.6 \times 10^{-3}$ &4.8(20)$\times 10^{-4}$ \cite{Kreek94}
			\\\noalign{\smallskip}

		\end{tabular}
	\end{ruledtabular}
	
	\label{rates}
\end{table}

In Fig.~\ref{PECDF}, the presently measured $P_{\rm ECDF}$ in $^{242}$Es and $^{238}$Bk are shown as a function of their neutron numbers. In addition, other known cases of $P_{\rm ECDF}$ values in Es, Bk and Am are shown. Typically, the $P_{\rm ECDF}$ are presented as a function of $Q_{\rm EC}$-$B_{\rm f}$, where $B_{\rm f}$ and $Q_{\rm EC}$ are the fission-barrier height of the even-even daughter nucleus and the $Q$ value of the EC-decaying odd-odd mother nucleus, respectively. In Fig.~\ref{PECDF}, these values, taken from the finite-range droplet model (FRDM, \cite{Moll95}) and finite-range liquid-drop model (FRLDM, \cite{Moll09}) are plotted as a function of neutron number. In these plots, one can clearly notice that the $P_{\rm ECDF}$ increases with an increase in $Q_{\rm EC}$-$B_{\rm f}$. However, the strength of such a relation is not invariant as a function of the proton number, $Z$. This can well be seen once we compare the $P_{\rm ECDF}$ values of Es, Bk and Am corresponding to $Q_{\rm EC}$-$B_{\rm f}=0$, as marked in Fig.~\ref{PECDF}. Experimentally known $P_{\rm ECDF}$ decreases with a decrease in $Z$, which indicates that there is a certain dependence between $P_{\rm ECDF}$ and $Z$ of the ECDF decaying-nucleus. 

Such a relation has recently been explained by Khuyagbaatar \cite{Khu19d}, via the effect of the fission-barrier shape of heavy nuclei. In this work, the probability of fission from excited states in even-even nuclei formed as daughters of EC-decaying odd-odd ones, has been described with two components: $P_{\rm f}=T\times P_{\rm R}$. The first term, $T$ is the probability for passage through a single-humped parabolic fission barrier, which has been calculated as a transmission coefficient by Hill and Wheeler \cite{HilW53a}. The second term, $P_{\rm R}$ was defined as a retardation factor for the fission, which accounts for the effect of the outer barrier. This factor was empirically extracted from the known cases of ECDF, but its absolute values were `normalized' to the experimental $P_{\rm ECDF}$ value in $^{248}$Es for which $P_{\rm R}$ was set to 1 \cite{Khu19d}. 
The results of the calculated $P_{\rm ECDF}$ values for Es, Bk and Am isotopes in Ref.~\cite{Khu19d} are also shown in Fig.~\ref{PECDF}. 

As one can see, the neutron-deficient Es isotopes with $N<148$ are well described with $P_{\rm R}=1$, which shows that the outer barrier remains ineffective for their DF probability. This is in fine agreement with the theoretically predicted smooth change in the barrier-shape for more neutron-deficient nuclei, i.e., in the outer barrier, and second well \cite{Moll09,Stas05,Khu20a}.
It is worth noting that the presently measured $P_{\rm R}$ for $^{242}$Es is in fine agreement with the prediction. In the cases of Bk and Am, the $P_{\rm ECDF}$ were well described with significantly reduced $P_{\rm R}$ values of $4\times 10^{-4}$ and $5\times 10^{-7}$, respectively. Such a decreasing trend has been attributed to the increasing effect of the outer barrier in the region of Cm and Pu isotopes, the barrier shapes of which are well-known to be of a multi-humped nature \cite{Moll09,BjoL80,Sobi07,Khu20a}. 

In this regard, the recently measured $P_{\rm ECDF}$ in $^{230}$Am \cite{Wilson17}, for which a lower limit was reported, is underestimated by the prediction. A reason for such a deviation may be associated with a change in the barrier shape. In other words, a ``sudden'' change in the barrier-shape for DF occurs in the neutron-deficient $^{230}$Pu
compared to heavier Pu isotopes. According to the calculations given in Ref.~\cite{Khu19d}, such a large experimental $P_{\rm ECDF}$ can be described with significantly increased $P_{\rm R}$, e.g., $10^{-4}$ as shown in Fig.~\ref{PECDF}. This shows the importance of measuring new cases of ECDF, which will be useful not only for verifying theoretically predicted $P_{\rm ECDF}$ values but also for probing the fission-barrier shapes of extremely neutron-deficient nuclei. 

Another interesting feature seen in Am isotopes is that even by taking larger $P_{\rm R}$, no saturation in $P_{\rm ECDF}$ values was predicted, in contrast to the cases of Es and Bk. 
This feature is due to the different behaviors of $B_{\rm f}$ and $Q_{\rm EC}$ values in the Am isotopes, which are shown in Fig.~\ref{PECDF}. In more neutron-deficient Am isotopes, the stabilization effect from the $N=126$ shell becomes significant in both $B_{\rm f}$ and $Q_{\rm EC}$. Especially, it has a large effect on fission, which results in an increase of $B_{\rm f}$ in more neutron-deficient nuclei. In turn, it reduces the fission probability, thus, the $P_{\rm ECDF}$. Therefore, despite the observed large $P_{\rm ECDF}$ in $^{230}$Am, it is still anticipated to observe non-saturated $P_{\rm ECDF}$ in more neutron-deficient, yet unknown Am isotopes with $N<130$.

\section{Summary and conclusion}

In the present work, the radioactive decay modes of the neutron-deficient isotope $^{242}$Es were studied with significantly larger statistics compared to previous studies. This isotope was produced with a cross section of 27(3)~nb in the fusion-evaporation reaction of $^{48}$Ca~+~$^{197}$Au. The current experimental results improve the relevant literature data, especially on the probability of the electron-capture delayed fission, which was measured to be 0.015(4). This, along with other known cases of ECDF in Es, Bk and Am isotopes were discussed within a recent semi-empirical approach, where the fission barrier shape was considered. The measurements of new and improved values for the ECDF probability in neutron-deficient heavy nuclei are necessary for further exploring the ECDF process. 

\section{Acknowledgments}

We are grateful to GSI's ion-source and UNILAC staff, the Experiment Electronics, and the Target Laboratory departments for their support of the experiment. The results presented here are based on the experiment U308, which was performed at the beam line X8/TASCA at the GSI Helmholtzzentrum f\"ur Schwerionenforschung, Darmstadt (Germany) in the frame of FAIR Phase-0.

%\nocite{*}
%\bibliographystyle{apsrev4-2}
\bibliography{main_242Es}% Produces the bibliography via BibTeX.

\end{document}